\documentstyle[preprint,aps,eqsecnum]{revtex}
\tightenlines

\def\m@thcombine#1#2{%
  \setbox0=\hbox{$#1$}
  \setbox1=\hbox{$#2$}
  \ifdim\wd0>\wd1
    \setbox0=\hbox to\wd1{\hss\box0\hss}
  \else
    \setbox1=\hbox to\wd0{\hss\box1\hss}
  \fi
  \mathop{\vcenter{
    \offinterlineskip\box0\box1}}}
\def\lesim{\m@thcombine<\sim}
\def\gesim{\m@thcombine>\sim}
\begin{document}

%%%%%%%%%%%%%%%%%%%%%%%%%%%%%%%%%%%%%%%%%%%%%%%%%%%%%%%%%%
\draft
\title{ TOPOLOGICAL STRUCTURE  OF CHIRAL QCD VACUUM }

\author{V. Gogohia$^{a,b}$ and H. Toki$^b$ }

\address{$^a$HAS, CRIP, RMKI, Depart. Theor. Phys., P.O.B. 49,
          H-1525, Budapest 114, Hungary \\
         $^b$Research Center for Nuclear Physics (RCNP), Osaka University \\
         Mihogaoka 10-1, Ibaraki, Osaka 567-0047, Japan \\ email addresses:
         gogohia@rcnp.osaka-u.ac.jp Cc: gogohia@rmki.kfki.hu}

\maketitle

\begin{abstract}
 Using the trace anomaly relation, low-energy theorem and
Witten-Veneziano formula, we have developed an analytical formalism which allows one to calculate the gluon condensate, the topological susceptibility and the
mass of the $\eta'$ meson in the chiral limit as functions of the non-perturbative vacuum energy density. It is used for numerical evaluation of the chiral 
QCD topology within the QCD vacuum model consisting mainly of the quantum component given by the recently 
proposed zero modes enhancement (ZME) model and the classical component given by the random instanton liquid model (RILM). We sum up both contributions
into the total, non-perturbative vacuum energy density. A very good agreement with the phenomenological values of the topological susceptibility, the mass of
the $\eta'$ meson in the chiral limit and the gluon condensate has been obtained. This puts the above mentioned QCD vacuum model on a firm phenomenological 
ground.
\end{abstract}


\vfill
\eject

\section{Introduction }

The nonperturbative QCD vacuum has a very rich dynamical and topological structure [1]. It is a very complicated medium and its dynamical and topological complexity means that its structure can be organized at various levels (quantum, classical) and it can contain many different components and ingredients which may
contribute to the vacuum energy density, the one of main characteristics of the
QCD ground state. The quantum part of the vacuum energy density is determined  
by the effective potential approach for composite operators introduced by      
Cornwall, Jackiw and Tomboulis (CJT) [2] (see also Ref. [3]). It allows us to investigate the non-perturbative QCD vacuum, since in the absence of external sources the effective potential is nothing but the vacuum energy density. It gives the vacuum energy density in the form of the loop expansion where the number 
of the vacuum loops (consisting of the confining quarks and non-perturbative gluons properly regularized with the help of ghosts) is equal to the power of the
Plank constant, $\hbar$.

 In our previous work [4] we have formulated
 a new, quantum model of the QCD ground state (its
non-perturbative vacuum), the so-called zero modes enhancement (ZME) model.
It is based on the existence and importance of such kind of the non-perturbative, topologically nontrivial quantum excitations of the gluon field configurations (due to the self-interactions of massless gluons only, i. e. without any extra degrees of freedom) which can be effectively, correctly described by the $q^{-4}$-type behaviour of the full gluon propagator in the deep infrared domain. 
The correct treatment of such strong singularity by the dymensional regularization method within the distribution theory was one of the highlights of the above mentioned preliminary publication [4]. Our model allows one to
calculate the non-perturbative vacuum energy density from first principles using the CJT approach for
 composite operators [2]. We have also formulated the method of how
 to determine numerically  the finite part of the vacuum energy density.
We propose to minimize the effective potential at a fixed scale as
function of a parameter which has a clear physical meaning. When
it is zero then only the perturbative phase remains in our model.
Equivalently one can minimize the corresponding auxiliary
effective potential as a function of the ultraviolet (UV) cut-off
itself. The non-perturbative chiral QCD vacuum is found stable
since its main characteristic -the vacuum energy density - has no
imaginary part and it is always negative.

Within the ZME quantum model of the QCD ground state [4],
the vacuum energy density depends on a scale at which the
 non-perturbative effects become
important. If QCD itself is confining theory, such a characteristic
 scale should certainly exist. The quark part of the vacuum energy density
  depends in addition on the constant of integration of the corresponding
 Schwinger-Dyson (SD) equation.
 The numerical value of the non-perturbative scale as
well as the above mentioned constant of integration is obtained
from the bounds
\begin{equation}
87.2 \leq  F^o_{\pi} \leq 93.3  \ (MeV),\end{equation}
for the pion decay constant in the chiral limit by implementing a
physically well-motivated scale-setting scheme [4]. We have
obtained the following numerical results for the non-perturbative
 vacuum energy density, $\epsilon = \epsilon_g + N_f \epsilon_q$,
\begin{eqnarray}
\epsilon = - (0.01425 - N_f 0.00196)  \ GeV^4, \\
\epsilon = - (0.01087 - N_f 0.00150)  \ GeV^4,
\end{eqnarray}
where, obviously, the first and second values are due to upper and lower bounds
in (1.1), respectively. Let us remind that these numerical values have been
 obtained by approximating the full gluon propagator by its non-perturbative
 term in the whole range, i. e. it has been already assumed that the perturbative contribution has been already subtracted.
 Let us recall that here and further on below $N_f$ is the number of light flavors and in what follows we will use $\epsilon = \epsilon_g \equiv \epsilon_{YM}$ in (1.2-1.3) in the case of pure Yang-Mills (YM) fields, $N_f=0$.

 On the other hand, many models of the QCD vacuum involve some extra classical 
color field 
configurations (such as randomly oriented domains of constant color magnetic 
fields, background gauge fields, averaged over spin and color, stochastic colored background fields, etc) and ingredients such as color-magnetic and Abelian-projected monopoles (see Refs. [1,5] and references therein).  The relevance of
center vortices to QCD by both lattice [6] and analytical method [7] was recently investigated as well. However, the most elaborated classical models are the random
and interacting instanton liquid models (RILM and IILM) of the QCD vacuum [8]. 
They are based on the existence of the topologically nontrivial instanton-type 
fluctuations of gluon fields, which are solutions to the classical equations of
motion in Euclidean space [8] (and references therein). 

In this paper we treat the chiral QCD vacuum as consisiting mainly of the two components, classical one given by RILM [8] and quantum one given by ZME [4] by 
summing up their contributions into the total, non-perturbative vacuum energy  
density. The main purpose of this paper is to show that this model of the
non-perturbative QCD vacuum is in fair agreement with phenomenology. For example, it exactly reproduces the phenomenological value of the topological susceptibility. In sections 2, 3 and 4 using the trace
anomaly relation [9], low-energy theorem [10,11] and Witten-Veneziano (WV) formula [12] we develop an analytical formalism which allows us to calculate the gluon condensate, the topological susceptibility and the mass of the $\eta{'}$ meson in the chiral limit as functions of the total, non-perturbative vacuum energy density (the bag
constant, apart from the sign, by definition). In section 5 we present our estimate of the non-perturbative vacuum energy density in the chiral limit due to instantons. Section 6 is devoted to discussion and our conclusions are given in 
section 7. The numerical results are shown in Tables 1-9.

\section{ The gluon condensate in the strong coupling limit}

The vacuum energy density is important in its own right as the
main characteristic of the non-perturbative vacuum of QCD.
Furthermore it assists in estimating such an important
phenomenological parameter as the gluon condensate, introduced in
the QCD sum rules approach to resonance physics [13]. The famous
trace anomaly relation [9] in the general case (nonzero current
quark masses $m_f^0$) is
\begin{equation}
\Theta_{\mu\mu} = {\beta(\alpha_s) \over 4 \alpha_s}
G^a_{\mu\nu} G^a_{\mu\nu}
+ \sum_f m_f^0 \overline q_f q_f.
\end{equation}
where $\Theta_{\mu\mu}$ is the trace of the energy-momentum tensor
and $G^a_{\mu\nu}$ being the gluon field strength tensor while $\alpha_s = g^2/4 \pi$.
Sandwiching (2.1) between vacuum states and on account of the
obvious relation $\langle{0} | \Theta_{\mu\mu} | {0}\rangle = 4
\epsilon_t$, one obtains
\begin{equation}
\epsilon_t = {1 \over 4}
\langle{0} | {\beta (\alpha_s)  \over 4 \alpha_s} G^a_{\mu\nu} G^a_{\mu\nu} |
 {0}\rangle
+ {1 \over 4} \sum_f m^0_f \langle{0} | \overline q_f q_f | {0}\rangle,
\end{equation}
where $\epsilon_t$ is the sum of all possible independent, non-perturbative
contributions to the
vacuum energy density (the total vacuum energy density) and $\langle{0} |
 \overline q_f q_f | {0}\rangle$ is the quark condensate.
  From this equation in the
chiral limit ($m^0_f=0$), one obtains
\begin{equation}
\langle \bar G^2 \rangle \equiv -
\langle {\beta(\alpha_s) \over 4 \alpha_s} G^2 \rangle \equiv -
\langle{0} | {\beta(\alpha_s) \over 4 \alpha_s} G^a_{\mu\nu} G^a_{\mu\nu}
 | {0}\rangle = - 4 \epsilon_t,
\end{equation}
where we need to introduce a new quantity, namely the gluon condensate in the strong coupling limit, i. e. not using in general the weak coupling limit solution to the $\beta$-function (see section 5 below).
If confinement happens then the $\beta$-function is
always in the domain of attraction (i. e. always negative) without IR stable fixed point [14]. Thus the non-perturbative gluon condensate $\langle
\bar G^2 \rangle$, defined in (2.3), is always positive as it should be.
Saturating $\epsilon_t$ by our values (1.2-1.3) which are relevant
in the strong coupling limit, one obtains

\begin{equation}
\bar G_2 \equiv \langle \bar G^2 \rangle = - 4 (\epsilon_g + N_f \epsilon_q),
\end{equation}
which gives the gluon condensate in the strong coupling limit as a
function of $N_f$. The numerical results are shown in Table 1.

\section{ The topological susceptibility}

One of the main characteristics of the QCD non-perturbative vacuum is
 the topological density operator (topological susceptibility)
  in gluodynamics ($N_f=0$) [12]

\begin{equation}
\chi_t =  \lim_{q \rightarrow 0} i \int d^4x\, e^{iqx} { 1 \over N_c^2} \langle{0} | T \Bigl\{ {q(x) q(0)} \Bigr\} | {0} \rangle,
\end{equation}
where $q(x)$ is the topological charge density, defined as
$q(x) = (\alpha_s / 4 \pi) F (x) \tilde{F} (x) = (\alpha_s / 4 \pi)
F^a_{\mu \nu} (x) \tilde {F}^a_{\mu \nu}(x)$
and $\tilde {F}^a_{\mu \nu}(x) = (1 / 2)
\epsilon^{\mu \nu \rho \sigma} F^a_{\rho \sigma} (x)$
is the dual gluon field strength tensor, while $N_c$ is the number of different
colors. In the definition of the topological
susceptibility (3.1) it is assumed that the corresponding regularization
and subtraction of the perturbative contribution have been already done in
order (3.1) to stand for the renormalized, finite and the non-perturbative
 topological susceptibility (see Refs. [10-12,15]).
Precisely this quantity measures the fluctuation of the
topological charge in the non-perturbative vacuum.

The anomaly equation in the WV notations is

\begin{equation}
\partial_{\mu} J_5^{\mu} = N_f (2/N_c) ({\alpha_s \over 4 \pi} ) F \tilde{F}.
\end{equation}

As it was shown in Refs. [10,11], the topological susceptibility
can be related to the non-perturbative gluon condensate via the
low energy theorem in gluodynamics as follows
\begin{equation}
\lim_{q \rightarrow 0} i \int d^4x\, e^{iqx} \langle{0} | T \Bigl\{ {\alpha_s
\over 8 \pi} G \tilde{G} (x)
{\alpha_s \over 8 \pi} G \tilde {G}(0) \Bigr\} | {0} \rangle = - \xi^2
\langle { \beta(\alpha_s) \over 4 \alpha_s} G^2 \rangle.
\end{equation}
There exist two proposals to fix the numerical value of the coefficient
$\xi$. The value $\xi = 2/b, \ b=11$ was suggested long time ago by
 Novikov, Schifman, Vanshtein and Zakharov (NSVZ), who used the dominance of
  self-dual fields hypothesis in the YM vacuum [10].
A second one, $\xi = 4/3b$, was advocated very recently by Halperin and
 Zhitnitsky (HZ), using a one-loop connection between the conformal and axial
  anomalies in the theory with auxiliary heavy fermions [11] (and references therein). However, in our numerical calculations we will use both values for
the $\xi$ parameter since the precise validity neither of the WV formula (4.3) 
below nor the NSVZ-HZ low energy relation (3.3) is not completely clear. 

The anomaly equation in the NSVZ-HZ notations is

\begin{equation}
\partial_{\mu} J_5^{\mu} = N_f ({\alpha_s \over 4 \pi} ) G \tilde{G},
\end{equation}
with $N_f=3$. Thus in order to get the topological susceptibility in the
WV form from the relation (3.3), it is necessary to make a replacement 
in its left hand side as follows: $G \tilde{G} \rightarrow (2/N_c) F \tilde{F}$
in accordance with anomaly equations (3.2) and (3.4). 
Then the WV topological susceptibility (3.1) finally becomes

\begin{equation}
\chi_t = - \xi^2 \langle { \beta(\alpha_s) \over 4 \alpha_s} G^2 \rangle
= - (2 \xi)^2 \epsilon_{YM},
\end{equation}
where the second equality comes from Eqs. (2.3) and (2.4) at $N_f=0$. The significance of this formula is that it gives the topological
 susceptibility as a function of the non-perturbative vacuum energy density for pure gluodynamics, $\epsilon_{YM}$.
                             
It is easy to show that one obtains the same expression for the topological susceptibility (3.5) as a function of the non-perturbative YM vacuum energy density if one would use the weak coupling limit solution to the $\beta$-function from the very beginning (see section 5 below). The numerical results 
due to ZME model are shown in Table 2.
In conclusion, let us note that there exists an obvious relation between the
HZ and NSVZ values for the $\xi$ parameter, namely $\xi_{HZ} =
(2/3) \xi_{NSVZ}$.

\section{ The $U(1)$ problem}

 The topological susceptibility (3.1) assists
in the resolution of the $U(1)$
problem\footnote{We are going to consider
 here only one aspect of this problem,
 namely the large mass of the $\eta'$ meson.}
[16] via the WV formula for the mass of the $\eta'$ meson [12].
 Within our notations it is expressed (in the chiral limit ) as follows:
$f^2_{\eta'} m^2_{\eta'} = 4 N_f \chi_t$,
where $f_{\eta'}$ is the $\eta'$ residue defined in general as
$\langle {0}| \sum_{q=u,d,s} \overline q \gamma_{\mu} \gamma_5 q | {\eta'}
\rangle = i \sqrt{N_f} f_{\eta'} p_{\mu}$ and $\langle {0}| N_f {\alpha_s \over 4 \pi} F \tilde{F} | {\eta'} \rangle = (N_c \sqrt{N_f} / 2) f_{\eta'} m^2_{\eta'}$ [12].
 Using also the normalization relation $f_{\eta'} = \sqrt{2} F^0_{\pi}$, one
 finally obtains

\begin{equation}
F^2_{\pi}m^2_{\eta'} = 2 N_f \chi_t.
\end{equation}
Then Eq. (3.5) implies

\begin{equation}
m^2_{\eta'} = - 2 N_f \Bigl( {2 \xi \over  F_{\pi}} \Bigr)^2 \epsilon_{YM}.
\end{equation}
In previous expressions we omit for simplicity the
superscript "0" in the pion decay constant as well as in
$m^2_{\eta'}$. In the numerical evaluation of the
 expression (4.2), we will put, of course, $N_f=N_c=3$, while the topological
 susceptibility will be evaluated at $N_f=0$ as it should be by definition.
This equation
 expresses the mass of the $\eta'$ meson as a function of the non-perturbative vacuum energy density which allows one to easily calculate it
in the chiral limit within our formalism (see again Table 2).

 It is instructive to reproduce the WV formula (4.1) in the non-chiral case as
well, namely
\begin{equation}
m^2_{\eta'} = {2 N_f \over F^2_{\pi}} \chi_t + \Delta ,
\end{equation}
where $\Delta = 2 m^2_{K} - m^2_{\eta}$.
 The precise validity of the WV formula
(4.3) is, of course, not completely clear, nevertheless, let us
regard it (for simplicity) as exact .
 Using now experimental values of all physical quantities
entering this formula, one obtains that the phenomenological
("experimental") value of the topological susceptibility is
\begin{equation}
\chi_t^{phen} = 0.001058 \ GeV^4 = (180.36 \ MeV )^4.
\end{equation}

In the chiral limit $\Delta = 0$ since $K^{\pm}$ and $\eta$ particles are
 Nambu-Goldstone (NG) bosons. Omitting formally this contribution from the
right hand side of Eq. (4.3) and on account of (1.1), one is able to derive an
upper and absolute lower bounds for the mass
of the $\eta'$ meson in the chiral limit

\begin{equation}
854 \leq m^0_{\eta'} \leq 913.77 (\ MeV),
\end{equation}
which should be compared with its experimental value
$m^{exp}_{\eta'}  = 957.77 \ MeV$. One can conclude that the mass of $\eta'$ meson remains large even in the chiral limit. It is worth noting that neither the numerical value of the topological susceptibility nor the mass of the $\eta'$ meson in the chiral limit can not exceed their phenomenological and experimental values. So the WV formula (4.3) in the chiral limit provides an absolute lower bound for the pion decay constant in this case, namely   $F^0_{\pi} \geq 83.2 \ MeV$.  

In order to directly apply this formalism to RILM we need the realistic estimate of the corresponding chiral vacuum energy density in this model.

\section{ The vacuum energy density due to instantons}

The instanton-type topological fluctuations, being
a classical phenomena, nevertheless also contributed to the vacuum energy
 density through a tunneling effect which was known to lower the energy of
the ground state [8]. It can be estimated 
as follows. Let us consider again the trace anomaly relation (2.2) in the chiral limit, i. e.
$\epsilon_t = (1/ 4)
\langle{0} | (\beta (\alpha_s) / 4 \alpha_s) G^a_{\mu\nu} G^a_{\mu\nu}
 | {0}\rangle$.
Using the weak coupling limit solution to the $\beta$-function now

\begin{equation}
\beta(\alpha_s) = - b { \alpha^2_s \over 2 \pi} + O(\alpha^3_s), \qquad
 b = 11 - {2 \over 3} N_f,
\end{equation}
one obtains

\begin{equation}
\epsilon_t = - {b \over 4} \times {1 \over 8}
\langle{0} | {\alpha_s \over \pi } G^a_{\mu\nu} G^a_{\mu\nu} | {0}\rangle.
\end{equation}
The phenomenological analysis of QCD sum rules [13] for the gluon condensate
 implies

\begin{equation}
G_2 \equiv \langle G^2 \rangle \equiv
 \langle {\alpha_s \over \pi} G^2 \rangle \equiv
\langle{0}|{\alpha_s \over
\pi}G^a_{\mu\nu}G^a_{\mu\nu}|{0}\rangle \simeq  0.012 \ GeV^4,
\end{equation}
which can be changed within a factor of two [13].
From the phenomenological  estimate (5.3), one easily can calculate
$(1 / 8) \langle{0} | (\alpha_s / \pi) G^a_{\mu\nu} G^a_{\mu\nu}
 | {0}\rangle \simeq  0.0015 \ GeV^4 \simeq  1.0 \ fm^{-4}$.
  Having in mind this and assuming that the gluon condensate in the weak
coupling limit is determined by the instanton-type fluctuations only, Shuryak
[8] (see also references therein) has concluded in that the "average separation" between instantons was $ R \simeq 1.0 \ fm$, so the corresponding density of
the instanton-type fluctuations should be $ n \simeq 1.0 \ fm^{-4}$. Let us note that the second parameter of the instanton liquid model of the QCD vacuum, the instanton size $\rho_0 \simeq 1/3$, was chosen to reproduce standard (also (as gluon condensate) phenomenologically estimated from QCD sum rules [13]) value
of the quark condensate. This contribution to the vacuum energy density via the
trace anomaly relation (2.1-2.2) vanishes in the chiral limit.  However, due to
all reasonable estimates of light quark masses, numerically its contribution  
is at $20\%$ and thus comparable with the systematic error in the determination
of the gluon condensate itself [13,17].  

Saturating the total vacuum energy in the weak coupling limit by instanton component $\epsilon_I$ and using the above mentioned estimate,
from (5.2) for dilute ensemble, one finally obtains

\begin{equation}
\epsilon_I = - {b \over 4} n = - {b \over 4} \times 1.0 \ fm^{-4}
= - (0.00417 - N_f 0.00025 ) \ GeV^4.
\end{equation}
Thus instanton contribution to the vacuum energy density was not calculated
independently but was postulated via the trace anomaly relation using the
phenomenological value of the gluon condensate (5.3) as well as weak coupling
limit solution to the $\beta$-function (5.1).
It is well known that density of instanton-type fluctuations is suppressed in the chiral limit and is again restored bacause of dynamical breakdown of chiral 
symmetry [8] (and references therein). In any case it
can not be large in the chiral limit, so the functional dependence of the vacuum energy density on the instanton density, established in Eq. (5.4) due to dilute gas approximation, seems to be justified in this
case. The only problem is the numerical value of the instanton density itself, 
which can be taken either from
phenomenology or from lattice simulations.

 In Ref. [10] it was argued that the gluon condensate in the chiral limit is
  approximately two times less than the above mentioned phenomenological
  (empirical) value (5.3), i. e.
   $\langle G^2 \rangle_{ch} \simeq 0.5 \langle G^2 \rangle_{phen}$.
This means that in this case instanton density $n \simeq 0.5 \ fm^{-4}$ and
the vacuum energy density due to instantons approximately two times less
than (5.4). However, it has been already pointed out [18] that QCD sum rules substantially underestimate the value of the gluon condensate.
The most recent phenomenological calculation of the gluon condensate is
given by Narison in Ref. [19], where a brief review of many previous calculations is also presented. His analysis leads to the update average value as

\begin{equation}
\langle{0}|{\alpha_s \over
\pi}G^a_{\mu\nu}G^a_{\mu\nu}|{0}\rangle = (0.0226 \pm 0.0029) \ GeV^4.
\end{equation}
This means that instanton density is approximately two time bigger than it was 
estimated by Shuryak for instanton liquid model [8], but in the chiral limit we
are again left with (5.4). 

In lattice QCD situation with instaton density and their sizes is also ambigious. In quenched ($N_f=0$) lattice QCD by using the so-called "cooling"
method the role of the instanton-type fluctuations in the QCD vacuum was
investigated [20]. In particular, it was found that the instanton density should be $n=(1+ \delta) \ fm^{-4}$, where $\delta \simeq 0.3-0.6$ depending on cooling steps. Moreover, by studying the topological content of the vacuum of $SU(2)$ pure gauge theory using a method of RG mapping [21], it is concluded that the average radius of an instanton is about $0.2 \ fm$, at a density
of about $2 \ fm^{-4}$. However, in Ref. [22] the topological content of the   
$SU(3)$ vacuum was studied using the same method as for $SU(2)$ gauge theory earlier and was obtained a fair agreement with Shuryak's phenomenologically estimated numbers for the instanton liquid model. At the same time, in Refs. [23,24]
considerably larger values were reported. Thus at this stage it is rather difficult to choose some well-justified numerical value of the instanton-type contribution to the non-perturbative vacuum energy density. In any case, in what follows we will consider (5.4) as a realistic upper bound for the instanton contribution to the vacuum energy density in the chiral limit. If the instanton number
density is about $n \simeq 2 \ fm^{-4}$, then in the chiral limit we again are 
left with (5.4), but if it is about $n \simeq 1 \ fm^{-4}$, we will be left with half of (5.4). Then the instanton contributions to the topological
susceptibility and the mass of the $\eta'$ meson in the chiral limit are to be
calculated via (3.4) and (4.2), respectively on account of the substitution
$\epsilon_{YM} \longrightarrow \epsilon_I (N_f=0)$, where $\epsilon_I$ is given
in (5.4) with the two different values for instanton number densities in the chiral limit, $n = 0.5 \ fm^{-4}, \ 1.0 \ fm^{-4}$. The numerical results are shown in Table 3. In conclusion, we note that for densities $n > 2 \ fm^{-4}$ 
(which means $n > 1 \ fm^{-4}$ in the chiral limit) the applicability of the dilute gas approximation becomes, apparently, doubtful.

\section{Discussion}

\subsection{The gluon condensate}

 It becomes almost obvious that we must  distinguish the two types of gluon
 condensates, both of which are the non-perturbative quantities. The first
  one is determined by (2.3) and is the one which is relevant in
the strong coupling limit. In this case the total vacuum energy is mainly
 saturated by the ZME component as it is precisely shown in (2.4). In the weak
  coupling limit, saturating $\epsilon_t$ by $\epsilon_I$, from (5.2-5.4)
  one obtains

\begin{equation}
 \langle {\alpha_s \over \pi} G^2 \rangle  =
- {32 \over b} \epsilon_t \simeq - {32 \over b} \epsilon_I = 8 n,
\end{equation}
i. e. gluon condensate in the weak coupling limit does not explicitly depend on
$N_f$. As was mentioned above, precisely this gluon condensate was introduced
long ago [13]. This unphysical situation takes place because in instanton
 calculus [8] there is no other way to calculate the vacuum energy density than the
trace anomaly relation (2.1-2.2) which becomes finally (6.1) as it
was described above. In this case it is preferable to have the
$N_f$ dependent vacuum energy density than the gluon condensate
since the former is the main characteristic of the
non-perturbative vacuum. Contrast to this, we have calculated the
vacuum energy density completely independently from the trace
anomaly relation.
We use it only to calculate the gluon condensate in the strong coupling limit.
That is why in our case both quantities are $N_f$ dependent functions.

Let us make a few things perfectly clear. It makes sense to underline once more
that the vacuum energy density is $not$ determined by the trace anomaly relation (2.2). The real situation is completely opposite. As it was already mentioned
in the Introduction, the QCD vacuum beeing a very complicated medium, may contain many different components and ingredients which contribute to the vacuum energy density. These contributions are completely independent from the gluon condensate, of course. For example, in the chiral limit the explicit quark contribution to the vacuum emergy density via the trace anomaly relation (2.2) vanishes
. However, because of the dynamical chiral symmetry breaking (DCSB), there is nonvanishing (even in the chiral limit) explicit contribution to the quantum    
part of the vacuum energy density which comes from the vacuum quark loops as it
was described in our work [4] in detail. Thus, the total vacuum energy density,
defined as 
the trace of the energy-momentum tensor, becomes the sum of all independent contributions. This sum precisely determines the realistic value of the gluon condensate in the chiral limit via the trace anomaly relation and not vice versa.  
The gluon condensate may exist or not, but the vacuum energy density always exists (at least its quantum part) due to non-perturbative solutions to quark and 
gluon SD equations by substituting them into the effective potential (see Ref. 
[4] again). In other words, the vacuum energy density is much more fundamental 
quantity than the gluon condensate.  
     
Our bounds for full QCD ($N_f=3$) gluon condensate in the strong coupling limit

\begin{equation}
0.025 \leq - \langle{0}|{\beta (\alpha_s) \over 4 \alpha_s }G^a_{\mu\nu}G^a_{\mu\nu}|{0}\rangle \leq 0.033 \ (GeV^4),
\end{equation}
are comparable with recent phenomenological determination of the standard gluon
condensate by Narison (5.5).
%\begin{equation}
%0.04 \leq \langle{0}|{\alpha_s \over
%\pi}G^a_{\mu\nu}G^a_{\mu\nu}|{0}\rangle \leq 0.105 (\ GeV^4),
%\end{equation}
%recently derived from the families of $J/ \Psi$ and $\Upsilon$ mesons in
%Ref. [27] are consistent with our results.                                   
The parameterization (the left hand side)
of the two types of the gluon condensate may be, of course, the same but their
numerical values (the right hand sides) are not to be the same.
This difference is not only due to different physical observables as was
noticed in Ref. [8]. Though both quantities are the non-perturbative phenomena,
nevertheless this difference reflects different underlying physics. Our gluon condensate (2.3) is the strong coupling limit result and reflects the
nontrivial topology of the true QCD vacuum where quantum excitations of
gluon fields play an important role.
As was shown in our preceding papers [4,25] precisely
these type of gluon field configurations are mainly responsible for quark
confinement and DCSB. At the same time, the standard gluon condensate (5.3) is
the weak coupling phenomenon due to classical instanton-type fluctuations
in the true QCD vacuum which by themselves do not confine quarks [21,26-28].

Concluding let us note that in the lattice simulations
there already exist calculations of the gluon condensate which are one order
of magnitude bigger than the standard value, namely $ G_2 \simeq
0.1046 \ GeV^4$ for $SU(3)$ in Ref. [29] and $ G_2 \simeq 0.1556 \ GeV^4$
for $SU(2)$ in Ref. [30] ( see also review [8]). In phenomenology also there exist large values, namely

\begin{equation}
0.04 \leq \langle{0}|{\alpha_s \over
\pi}G^a_{\mu\nu}G^a_{\mu\nu}|{0}\rangle \leq 0.105 (\ GeV^4),
\end{equation}
which were recently derived from the families of $J/ \Psi$ and $\Upsilon$ mesons in Ref. [31].

\subsection{Topology of chiral QCD vacuum}

Our numerical results for the quantum part of the non-perturbative vacuum energy density and for the topological susceptibility with the mass of the $\eta'$meson in the chiral limit are presented
in Eqs. (1.2-1.3) and in Table 2, respectively. In general our values for the 
vacuum energy density are an order of magnitude bigger than RILM can provide at all in various modifications (see Table 3). That is why the quantum part of 
the non-perturbative vacuum energy density saturates the phenomenological value
of the topological susceptibility and the mass of the $\eta'$ meson in the chiral limit much better than the classical part given by instantons (compare Tables 2 and 3). Especially this is obvious for the HZ value of the $\xi$ parameter,
introduced in the low-energy theorem, Eq. (3.3).
The instanton contribution substantially underestimates the phenomenological value of the topological susceptibility and therefore can not account for the large mass of the $\eta'$ meson in the chiral limit alone (see Table 3).       

However, the total vacuum energy density, $\epsilon_t$, is, in principle, the 
sum of all possible independent, the non-perturbative contributions. Thus, at 
least it is the sum of the two well-established contributions, quantum
$\epsilon \equiv \epsilon_{ZME}$ and classical
$\epsilon_I$, i. e. $\epsilon_t = \epsilon_{ZME} + \epsilon_I + ...$,
where the dots denote other possible independent
contributions. In this case an excellent agreement with phenomenology
is achieved indeed (see Tables 4 and 5). The numerical values of the bag
constant $B$, defined as the difference between the perturbative and
non-perturbative vacua are given now by the relation $B = - \epsilon_t$ and can be explicitly evaluated using Eqs. (1.2-1.3) for $\epsilon_{ZME}$ and Eq. (5.4) for $\epsilon_I$ on account of the above mentioned two different instanton number densities (see Tables 6-9). For the readers convenience the bag constant (and consequently the total, non-perturbative vacuum energy density) is given in
often used different physical units.

\section{Conclusions}

In summary, using the trace anomaly relation, NSVZ and HZ low-energy theorem and Witten-Veneziano formula, we have developed an analytical formalism which allows one to calculate the gluon condensate, the topological susceptibility and  
the mass of the $\eta'$ meson in the chiral limit as functions of the non-perturbative vacuum energy density.  
It was immediately used for numerical investigation of the chiral QCD
non-perturbative vacuum topology within the recently proposed
ZME quantum model. 
 We have explicitly shown that precisely our values for the
non-perturbative vacuum energy density (1.2-1.3) are of the
necessary order of magnitude in order to saturate the large mass of the $\eta'$
meson in the chiral limit. We have 
obtained good approximation to the phenomenological value of the topological
susceptibility as well. The HZ value of the 
$\xi$ parameter, introduced in the low-energy theorem (3.3), especially nicely 
saturates them (for all result mentioned above see Table 2). At the same time, 
it is clear that instanton-induced contribution should be added to our values in order to achive an excellent agreement with phenomenology. Indeed, from Table
5 it follows

\begin{equation}
\chi_t = (180.3 \ MeV )^4
\end{equation}
and consequently (as it should be)

\begin{equation}
m^0_{\eta'} = 854 \ MeV,
\end{equation}
which are in fair agreement with (4.4) and lower bound in (4.5), respectively.
Let us remind that these numbers have been obtained when the pion decay constant was precisely approximated by its experimental value. The above displayed
excellent agreement with phenomenological values of the corresponding quantities is achieved by summing up our contribution and instanton-induced
contribution into the total, non-perturbative vacuum energy density, i. e. the 
summation was done purely phenomenologically by simply summing up
the two well-established contributions (classical instanton's and quantum ZME's). How to take into account 't Hooft's instanton-induced interaction [32] at the fundamental quark level within our approach is not completely clear for us, though see paper [33] (and references therein).

Let us also make a few things perfectly clear. At low energies QCD is
governed by $SU_L(N_f) \times SU_R(N_f)$ chiral symmetry and its dynamical breakdown in the vacuum to the corresponding vectorial subgroup [34]. The chiral 
limit is not physical one but nevertheless remains very important theoretical 
limit since to understand the chiral limit physics means to correctly understand the dynamical structure of low-energy QCD as well as the topological properties of its ground state. So a realistic calculation of various physical quantities as
well as chiral properties of its vacuum becomes important. In particular, any 
model of the QCD vacuum should pass the chiral limit test in order to be justified for further extrapolation to the realistic (nonchiral) case. In our previous publication [4] ZME quantum model was formulated. Here we have explicitly 
shown its important and novel feature, namely it itself passes the chiral limit
test justified thereby for use in the nonchiral case as well. Complemented
by instanton-induced contribution it is in a fair agreement with phenomenology.
 
In conclusion a few remarks are in order. It is well-known that instanton-type
fluctuations require topological charge to be integer ($\pm 1$) and the vacuum 
angle, $\theta$ - nonzero, which violates $P$ and $CP$ invariance of strong
interactions [10,16]. The non-perturbative $q^{-4}$-type quantum excitations
do not require the introduction of the vacuum angle, $\theta$, at all.
It is quite possible that  topological charge in this case is
not restricted to integer values. It has been explicitly shown that fractional 
(non-integer) topological charge configurations are required to resolve the $U(1)$ problem [16,35].
However, the  $\theta$ dependence of the QCD non-perturbative
vacuum energy remains an important problem. For recent
developments of this problem in the large $N_c$ limit of
four-dimensional gauge theories see papers [36]. In particular, in
Refs. [37,38] it has been discussed that the picture of its
dependence in QCD for finite  $N_c$ might be more complicated than
that predicted by the large $N_c$ values.                                      

And finally, let us emphasize once more indisputable simplicity of our analytical calculation of the topological susceptibility (7.1) in comparison with indisputable complexity of its calculation by lattice method [21-24,30,39-41]. 

\acknowledgements

  One of the authors (V.G.) is grateful to the late Prof. V.N. Gribov for many
interesting discussions and remarks on non-perturbative QCD. He also would like
to thank I. Halperin for correspondence and A. Vainshtein for useful remarks.

\vfill

\eject


\begin{references}
\bibitem{1}
   Confinement, Duality and Nonperturbative Aspects of QCD (ed. by P. van Baal)
      NATO ASI Series B: Physics, vol.368; 
   Non-Perturbative QCD, Structure of the QCD Vacuum, (ed. by K-I.Aoki,
    O.Miymura and T.Suzuki), Prog.Theor.Phys.Suppl., {\bf 131} (1998);
   V.N.Gribov, LUND preprint, LU TP 91-7, (1991) 1-51
\bibitem{2}
   J.M.Cornwall, R.Jackiw and E.Tomboulis, Phys.Rev.,
    {\bf D10} (1974) 2428
\bibitem{3}
   A.Barducci et al, Phys.Rev., {\bf D38} (1988) 238; \\
   R.W.Haymaker, Riv.Nuovo Cim., {\bf 14} (1991) 1-89; \\
   K.Higashijima, Prog.Theor.Phys.Suppl., {\bf 104} (1991) 1-69
\bibitem{4}
   V.Gogohia, Gy.Kluge, H.Toki, T.Sakai, Phys.Lett., {\bf B453} (1999) 281;\\
   V.Gogohia, H.Toki, T.Sakai and Gy.Kluge, hep-ph/9810516
\bibitem{5}
   J.Hosek and G.Ripka, Z.Phys., {\bf A354} (1996) 177
\bibitem{6}
   P.de Forcrand and M. D'Elia, Phys.Rev.Lett., {\bf 82} (1999) 4582
\bibitem{7}
   J.M.Cornwall, Phys.Rev., {\bf D59} (1999) 125015
\bibitem{8}
   T.Schafer and E.V.Shuryak, Rev.Mod.Phys., {\bf 70} (1998) 323
\bibitem{9}
   R.J.Crewther, Phys.Rev.Lett., {\bf 28} (1972) 1421; \\
   M.S.Chanowitz and J.Ellis, Phys.Rev., {\bf D7} (1973) 2490; \\
   J.C.Collins, A.Duncan and S.D.Joglecar, Phys.Rev.,
    {\bf D16} (1977) 438
\bibitem{10}
   V.A.Novikov, M.A.Shifman, A.I.Vainshtein, V.I.Zakharov,
    Nucl.Phys., {\bf B191} (1981) 301
\bibitem{11}
   I.Halperin and A.Zhitnitsky,  Nucl.Phys., {\bf B539} (1999) 166
\bibitem{12}
   E.Witten, Nucl.Phys., {\bf B156} (1979) 269; \\
   G.Veneziano, Nucl.Phys., {\bf B159} (1979) 213
\bibitem{13}
   M.A.Shifman, A.I.Vainshtein and V.I.Zakharov, Nucl.Phys.,
   {\bf B147} (1979) 385, 448
\bibitem{14}
   W.Marciano and H.Pagels, Phys.Rep., {\bf C36} (1978) 1                      
\bibitem{15}
   G.Gabadadze, Phys.Rev., {\bf D58} (1998) 094015
\bibitem{16}
   G.A.Christos, Phys.Rep., {\bf 116} (1984) 251
\bibitem{17}
   M.Schaden, Phys.Rev., {\bf D58} (1998) 025016
\bibitem{18}
   J.S.Bell and R.A.Bertlmann, Nucl.Phys., {\bf B177} (1981) 218; \\
   R.A.Bertlmann et al., Z.Phys. C  {\bf 39} (1988) 231
\bibitem{19}
   S.Narison, Phys.Lett., {\bf B387} (1996) 162
\bibitem{20}
   M.-C.Chu, J.M.Grandy, S.Huang and J.W.Negele, Phys.Rev., {\bf D49} (1994)
   6039
\bibitem{21}
   T.DeGrand, A.Hasenfratz and T.G.Kovacs, Nucl.Phys., {\bf B505} (1997) 417   
\bibitem{22}
   A.Hasenfratz and C.Nieter, Nucl.Phys., (Proc. Suppl.) {\bf B73} (1999) 503
\bibitem{23}
   D.A.Smith and M.J.Teper, Phys.Rev., {\bf D58} (1998) 014505
\bibitem{24}
   P.de Forcrand et al., Nucl.Phys., B (Proc. Suppl.) {\bf 63A-C} (1998) 549;\\
   P.de Forcrand et al., hep-lat/9802017
\bibitem{25}
   V.Gogohia, Gy.Kluge and M.Prisznyak, Phys.Lett., {\bf B368} (1996) 221
\bibitem{26}
   S.Coleman in: The Whys of Subnuclear Physics, Intern. School of Subnuclear
   Physics, Erice, 1977, ed. by A.Zichichi (Plenum Press, NY, 1979) p.805
\bibitem{27}
   R.C.Brower, D.Chen, J.W.Negele and E.Shuryak, hep-lat/9809091; \\
   J.W.Negele, hep-lat/9810053
\bibitem{28}
   C.G.Callan, Jr., R.Dashen, D.J.Gross, F.Wilczek and A.Zee, Phys.Rev.,
    {\bf D18} (1978) 4684
\bibitem{29}
   M.Campostrini, A.Di Giacomo and Y.Gunduc, Phys.Lett., {\bf B225} (1989) 393
\bibitem{30}
   B.Alles, M.Campostrini, A.Di Giacomo, Y.Gunduc and E.Vicari, Phys.Rev.,
    {\bf D48} (1993) 2284
\bibitem{31}
   B.V.Geskenbein, V.L.Morgunov, Russ.Jour.Nucl.Phys., {\bf 58} (1995) 1873
\bibitem{32}
   G. 't Hooft, Phys.Rev., {\bf D14} (1976) 3432
\bibitem{33}
   E.Klempt, B.C.Metsch, C.R.Munz and H.R.Petry, Phys.Lett.,
    {\bf B361} (1995) 160
\bibitem{34}
   H.Pagels, Phys.Rep., {\bf C16} (1975) 219; \\
   J.Gasser and H.Leutwyler, Phys.Rep., {\bf 87} (1982) 77
\bibitem{35}
   R.J.Crewther, Nucl.Phys., {\bf B209} (1982) 413
\bibitem{36}
   E.Witten, Phys.Rev.Lett., {\bf 81} (1998) 2862; \\
   M.Shifman, Phys.Rev., {\bf D59} (1998) 021501
\bibitem{37}
   I.Halperin and A.Zhitnitsky,  Phys.Lett., {\bf B440} (1998) 77; \\
   I.Halperin and A.Zhitnitsky,  Phys.Rev.Lett., {\bf 81} (1998) 4071
\bibitem{38}
   G.Gabadadze, hep-th/9902191
\bibitem{39}
   E.Meggiolaro, Phys.Rev., {\bf D58} (1998) 085002
\bibitem{40}
   B.Alles, M.D'Elia and A.Di Giacomo, Nucl.Phys., {\bf B494} (1997) 281
\bibitem{41}
   R.Narayanan and R.L.Singleton Jr., Nucl.Phys., B (Proc. Suppl.) {\bf 63A-C}    (1998) 555
\end{references}
\end{document}